\documentclass[a4paper,12pt]{article}
\vbadness=\maxdimen

\usepackage[english]{babel} 
\usepackage[T1]{fontenc}
\usepackage[ansinew]{inputenc}
\usepackage{lmodern} 
\usepackage{graphicx} 
\usepackage{caption} 
\usepackage{subcaption} 
\usepackage{microtype}

\usepackage{amsmath}
\usepackage{amsthm}
\usepackage{amsfonts}
\usepackage{amssymb}
\usepackage{psfrag}

\usepackage[numbers, round, comma, sort&compress, authoryear]{natbib} 
\usepackage{here}
\usepackage{authblk}
\usepackage{cleveref}

\usepackage{setspace}

\def \IR{\hbox{{\rm I}\kern-.2em\hbox{{\rm R}}}}
\def \IF{\hbox{{\rm I}\kern-.2em\hbox{{\rm F}}}}

\begin{document}
\title{Inhibitory geostatistical designs for spatial prediction taking account of uncertain covariance structure}

\clearpage
\pagestyle{empty}
\newpage
\author{ Michael G. Chipeta$^{1,2,3}$, Dianne J. Terlouw$^{2, 3, 4}$, Kamija S. Phiri$^2$,\\  Peter J. Diggle$^1$ } 
\affil{$^1$Lancaster Medical School, Lancaster University, Lancaster, UK.\\
$^2$College of Medicine, University of Malawi, Blantyre, Malawi.\\
$^3$Malawi-Liverpool Wellcome Trust, Blantyre, Malawi.\\
$^4$Liverpool School of Tropical Medicine, Liverpool, UK.}
\maketitle

\pagestyle{plain} 
\setcounter{page}{1}
\pagenumbering{arabic}

{\bf Abstract}\\
The problem of choosing spatial sampling designs for investigating unobserved spatial phenomenon $S$ arises in many contexts, for example in identifying households to select for a prevalence survey to study disease burden and heterogeneity in a study region ${\cal D}$. We studied randomised inhibitory spatial sampling designs to address the problem of spatial prediction whilst taking account of the need to estimate covariance structure. Two specific classes of design are \textit{inhibitory designs} and \textit{inhibitory designs plus close pairs}. In an inhibitory design, any pair of sample locations must be separated by at least an inhibition distance $\delta$. In an inhibitory plus close pairs design, $n - k$ sample locations in an inhibitory design with inhibition distance $\delta$ are augmented by $k$ locations each positioned close to one of the randomly selected $n - k$ locations in the inhibitory design, uniformly distributed within a disc of radius $\zeta$. We present simulation results for the Mat\'{e}rn class of covariance structures. When the nugget variance is non-negligible, inhibitory plus close pairs designs demonstrate improved predictive efficiency over designs without close pairs. We illustrate how these findings can be applied to the design of a rolling Malaria Indicator Survey that forms part of an ongoing large-scale, five-year malaria transmission reduction project in Malawi.


\newpage
\section{Introduction} \label{sec:intro}

Geostatistics  is concerned with investigation of an unobserved  spatial  phenomenon $S = \{S(x): x \in {\cal D} \subset \IR^2\}$, where ${\cal D}$ is a geographical region of interest. Its particular focus is on investigations in which  the available data consist of measurements $y_i$  at a finite set of  locations $x_i \in {\cal D}$. Typically, each  $y_i$   can  be  regarded  as  a  noisy  version  of $S(x_i)$.  We  write  ${\cal X}  = \{x_1 , ..., x_n\}$ and call ${\cal X}$ the {\it sampling  design}. Geostatistical {\it analysis}  mainly addresses two broad scientific objectives: {\it estimation} of the  parameters that define a stochastic model  for the unobserved process $S$  and  the  observed data  $Y = \{(y_i , x_i ) : i = 1, ..., n\}$; {\it prediction}  of the  unobserved  realisation  of $S(x)$ throughout ${\cal D}$, or particular characteristics of this realisation, for example its average  value. The fundamental geostatistical {\it design}  problem is the specification of ${\cal X}$. A key consideration is that sampling designs that are efficient for parameter estimation may be inefficient for prediction, and vice versa \citep{Zimmerman2006}. In practice, most geostatistical problems focus on spatial prediction, but parameter estimation is an important means to this end. Hence, there is a need to compromise between designing for efficient parameter estimation and designing for efficient prediction given  the values of relevant model parameters. In practice, selection of covariates and estimating their effects are also important considerations for study design. However, in this paper we focus on the design implications of the spatial covariance structure of $S$, this being the distinguishing feature of geostatistical, as opposed to general statistical, methodology.

In a previous paper \citep{Chipeta2016}, we have discussed {\it adaptive} geostatistical designs, in which sampling locations are chosen sequentially, either singly or in batches, and at any stage the analysis of already collected data can inform the selection  of the next batch of locations. In this paper, we consider {\it non-adaptive} geostatistical designs, in which  the complete design  ${\cal X}$ must be chosen in advance of any data-collection.

Two examples of non-adaptive designs are {\it completely random} and {\it lattice} designs. In a completely random design the locations $x_i$ are an independent random sample from the  uniform distribution on ${\cal D}$. In a lattice design, the $x_i$ form a regular (typically square) lattice to cover ${\cal D}$.  A combination of theoretical and empirical work, from Mat\'{e}rn (1960) onwards, has led to general acceptance  that lattice designs should  lead to efficient spatial prediction provided model parameters are known.  If model parameters are unknown, a completely random design has the advantage that it will include a wider range of inter-point distances, and in particular some small inter-point distances, and so provides more information on the shape of the covariance function of $S$. \citet{Diggle2006} described and compared empirically some compromise designs. In their simulations,  a lattice design supplemented by some close pairs of points performed well.

A limitation of lattice-based designs is that their absence of a probability sampling frame  leaves open the possibility of systematic bias. In the present paper we therefore propose a class of randomised {\it inhibitory plus close pairs}  designs to address the problem of spatial prediction whilst taking account of the need to estimate spatial covariance structure. We evaluate the performance of this class of designs through simulation studies and describe an application to data from a malaria transmission reduction monitoring and evaluation study in the Chikwawa district of southern Malawi. 

In \Cref{sec:design-strategies} we review the existing literature on non-adaptive geostatistical design strategies. In \Cref{sec:inhibitory} we describe our proposed class of designs. \Cref{sec:simulate} reports on simulation study of the predictive performance of the proposed design class. \Cref{sec:applic} describes an application to the sampling design of an ongoing malaria prevalence mapping exercise around the perimeter of the Majete wildlife reserve, Chikwawa district, Malawi. \Cref{sec:discuss} is a concluding discussion. All computations for the paper were run on the High End Computing Cluster at Lancaster University, using the \textbf{R} software environment \citep{RcoreTeam2015}.

\section{Non-adaptive geostatistical design strategies}\label{sec:design-strategies}

Different scientific goals and study settings require different geostatistical design strategies. Ideally, a design ${\cal X}$  will be chosen to maximise or minimise a performance criterion that reflects the primary objective of the study \citep{Jardim2007a, Nowak2010}. For example, a possible design criterion when the objective is to predict the value of $S(x)$ throughout the region ${\cal D}$ is the  spatially averaged mean squared  prediction error,

 \begin{equation}
 MSPE = \int_D {\rm E}[\{\hat{S}(x) - S(x)\}^2 dx,
 \label{eq:MSPE}
 \end{equation}

where $\hat{S}(x) = {\rm E}[S(x)|Y; {\cal X}]$ is the minimum mean square error predictor of $S(x)$ and expectations are with respect to  $S$. In practice, any such criterion needs to be tempered by application-specific  considerations of some kind, for example different costs and benefits of obtaining data and predictions, respectively, at particular locations.  

We review the following strategies for geostatistical designs: designing for efficient parameter estimation; designing for efficient spatial prediction when the covariance function is assumed completely known; and designing for efficient spatial prediction when the covariance function is not known and has to be estimated from the same data. \citet[Chapters~5 -- 7]{Muller2007} is a relatively recent book length account of geostatistical design strategies. 

Much of the work on spatial sampling design for estimating covariance structures has focused on estimation procedures based on the empirical variogram \citep{Muller1999, Russo1984, Warrick1987}.
\citet{Lark2002}  used likelihood estimation procedures under an assumed Gaussian process model.   \citet{Pettitt1993} studied several sampling designs for estimating parameters using the restricted maximum likelihood (REML) method of parameter estimation. A general consensus from this body of work is that completely random designs are efficient for parameter estimation. However, these designs have often been criticised because they  leave large unsampled ``swaths'' in the study region ${\cal D}$ \citep{Muller2007}.

Studies of design for efficient spatial prediction with known covariance structure include \citet{McBratney1981, McBratney1981b, Yfantis1987, Ritter1996, Su1993}. Spatially regular lattice designs, which achieve an even coverage of  ${\cal D}$, have been shown to be optimal in this case. Several variants of these designs have also been proposed. We provide definitions and an overview in \Cref{sec:design-classes}.

The assumption of a known covariance function is in most cases unrealistic \citep{Muller2007}. Usually we have to use the same data for estimation of covariance parameters and for spatial prediction, and effective prediction requires good estimates of the second order characteristics \citep{Guttorp1994}. Recent work on construction of designs that focus on the goals of efficient spatial prediction
in conjunction with parameter estimation includes \citet{Zhu2002, Zhu2006, Diggle2006, Pilz2006, Bijleveld2012, Banerjee2008} and \citet{Chipeta2016}.

\subsection{Classes of non-adaptive geostatistical designs}\label{sec:design-classes}

We now review several design classes that have been used for different analysis objectives: parameter estimation; spatial prediction; and a combination of the two. Design performance is largely influenced by \textit{sample pattern} and \textit{sample density} \citep{Olea1984}. `\textit{Pattern}' here refers to the geometrical configuration of sample points in a  given region, ${\cal D}$. `\textit{Density}' refers to the number of sample points per unit area. Both model-dependent and geometry-dependent designs have been proposed.

\subsubsection{Completely randomised designs}
In a  completely randomised design, locations $x_i, \; i = 1, \ldots, n$ are chosen independently, each with a uniform distribution over ${\cal D}$. This ensures that the design is
stochastically independent of the underlying spatial phenomenon of interest $S(x)$, which is a requirement for the validity of standard geostatistical inference methods \citep{Diggle2010d}. However, the resulting uneven coverage of ${\cal D}$ has a negative impact on spatial prediction. Variants of the completely random design include stratified and cluster random sampling \citep{Cressie1991}. These design strategies are well established in classical survey sampling; see, for example, \citet{Cochran1977}.

\subsubsection{Completely regular lattice designs}
Design points in this class form a regular lattice pattern over the study region ${\cal D}$, thereby ensuring an even coverage. The origin of the lattice should strictly be located at random \citep{Diggle2007}, although in practice this is often ignored. These designs are easy to implement and provide well defined directional classes within which variograms can be computed. Regular designs also have the potential of yielding computational savings over irregular designs such as those resulting from random sampling \citep{Cressie1991}. Regular lattice designs can use square, equilateral triangular or hexagonal grids. A comparison of the three suggests that the equilateral triangular grid design is the most efficient \citep{McBratney1981, McBratney1981b, Olea1984, Yfantis1987}. However, square lattices are more common in practice.

\subsubsection{Modified regular lattice designs}
\citet{Diggle2006} proposed and developed two different two-step  (augmented lattice) designs. These designs supplement a lattice with closely spaced pairs of points which, as noted earlier, are important for estimating certain parameters of the underlying spatial covariance structure, especially when this includes a nugget variance \cite[Chapter~8]{Diggle2007} or a smoothness parameter such as the shape parameter of a Mat\'{e}rn correlation function \citep{Zhu2006}.

Lattice plus close pairs design class is initialised with an even coverage of locations  in  ${\cal D}$; forming a $k \times k$ regular lattice at spacing $\Delta$. The design points are then augmented by a further location distributed uniformly at \emph{random} within a disc of radius $\delta = \alpha\Delta$ centred on each of the $m$ randomly selected lattice locations. Lattice plus infill design class is again initialised with an even coverage of $k \times k$ regular lattice at spacing $\Delta$, but is augmented with further locations in a more finely spaced lattice within 
$m$ randomly selected primary lattice cells. 

\subsubsection{Geometric designs}
\citet{Royle1998} describe a purely geometric design criterion for spatial prediction. This approach, commonly known as `space-filling' design, identifies sample locations by minimising a criterion that favours more regular geometrical configurations of sample locations \citep{Nychka1998}.

\subsubsection{Summary}	
Some general conclusions are the following. Good spatial prediction favours more regular than completely random designs when model parameters are known. When the analysis objective is parameter estimation, designs with a random configuration of design points are preferable. These two points suggest that some compromise is therefore needed when constructing designs for spatial prediction when model parameters have to be estimated from the same data. 

A good geostatistical design strategy also needs to be able to deal with a range of practical constraints. For example, potential sampling points may be limited to a finite set. This holds, for example, in our application to malaria monitoring, where data can only be collected from existing houses, within the study region.

\section{Inhibitory geostatistical designs}\label{sec:inhibitory}
\subsection{Design criterion}
We propose a class of \emph{inhibitory} geostatistical designs for spatial prediction when model parameters need to be estimated. We use $[\cdot]$ to mean ``the distribution of'' and incorporate a stochastic process $S = \{S(x): x \in \IR^2\}$ into a statistical model $[S, Y] = [S][Y|S]$, where $Y = (Y_1, \ldots, Y_n)$ are the  measured data-values at the points of ${\cal X}$ and $S = \{S(x_1), \ldots, S(x_n)\}$. The distribution for predictive inference is then the conditional distribution, $[S|Y]$, which follows from an application of Bayes' theorem as
	
	\begin{equation}\label{eq:stochastic-process}
		[S|Y] = [S][Y|S]/\int [S][Y|S]\mathrm{d}S
	\end{equation}
	
A typical spatial prediction problem involves making inferences about a functional $T = T(\{S(x): x \in {\cal D}\})$ given data $(Y_i,X_i), \;\; i = 1, \ldots, n$. In what follows, we use as performance criterion the average prediction variance, 
	
	\begin{equation}
		APV = \int_{\cal D} {\rm Var}\{S(x)|Y\}\mathrm{d}x
		\label{eq:apv}
	\end{equation}

\subsection{Simple inhibitory designs}\label{class1}	
An inhibitory design consists of $n$ locations chosen at random in ${\cal D}$ but with the constraint that no two locations are at a distance of less than some value $\delta$. Formally, the resulting design ${\cal X}$ is a realisation of a simple inhibitory point process that is itself a special case of a pairwise interaction point process; see, for example, \citet[Chapter~6]{Diggle2013}. This construction respects the established principles of random sampling theory, while guaranteeing some degree of spatial regularity. All designs ${\cal X}$ that meet the inhibitory constraint are equally likely to be picked. Also, the construction can be applied whether or not the potential sampling locations are confined to a finite set of points, although in either case the value of $\delta$ will limit the maximum achievable sample size.

We define the ``\textit{packing density}'' of the design to be the proportion of the total region covered by $n$ non-overlapping discs of diameter $\delta$, hence $\rho = (n\pi\delta^2)/(4|{\cal D}|)$. We use the notation \textbf{SI($n, \delta$)} and compare the performance of designs with fixed $n$ fixed and varying $\delta$. The formal construction of an \textbf{SI}$(n,\delta)$ design on a region ${\cal D}$ 
proceeds as follows:
\begin{enumerate}
	\item Draw a sample of locations $x_i : i = 1, \ldots, n$ completely at random in ${\cal D}$;
	\item Set $i = 1$;
  \item Calculate the minimum, $d_{min}$, of the distances from $x_i$ to all other $x_j$ in the current sample;
	\item If $d_{min} \geq \delta$, increase $i$ by 1 and return to step 3 if $i \leq n$, otherwise stop;
	\item If $d_{min} < \delta$, replace $x_i$ by a new location drawn completely at random in ${\cal D}$ and return to step 4.
\end{enumerate}

\subsection{Inhibitory design with close pairs}\label{class2}	
This class is defined by four scalars, namely: $n$, the total number of points; $\delta$, the minimum  distance between any two locations;  $k$, the number of close pairs and $\zeta$, the radius of the disc from the primary point within which to add a paired point. For a total of $n$ points, this design consists of $n - k$ points in an inhibitory design with inhibition distance $\delta$, augmented by $k$ points each positioned relative to one of the randomly selected $n - k$ points in the inhibitory design according to the uniform distribution over a disc of radius $\zeta$. We use the notation \textbf{ICP($n, k, \delta, \zeta$)}. The formal construction of an \textbf{ICP($n, k, \delta, \zeta$)} design on a region ${\cal D}$ proceeds as follows:
\begin{enumerate}
	\item Construct a simple inhibitory design \textbf{SI($n-k, \;\; \delta$)};
	\item Sample $k$ from $x_{1}, \ldots, x_{n-k}$ without replacement and call this set of locations $x^{*}_{j}, \;\; j = 1, \ldots, k$;
	\item For $j = 1, \ldots, k$, $x_{n - k + j}$ is uniformly distributed on the disc with centre $x^{*}_{j}$ and radius $\zeta$.
\end{enumerate}

Note that in the \textbf{ICP($n, k, \delta, \zeta$)} design, $k$ must be less than or equal to $n/2$. Also, when comparing an \textbf{SI($n, \delta$)} design with one or more \textbf{ICP($n, k, \delta, \zeta$)} designs, it is appropriate to require all of the inhibitory components have the same degree of spatial regularity. This requires $\delta$ to become a function of $k$, namely

\begin{equation}
\delta_{(k)} = \delta_0 \sqrt{n/(n - k)},
\label{eqn:delta}
\end{equation}

with $\delta_0$ held fixed. For fixed $n$, the minimum spacing between any two  inhibitory points therefore increases with $k$. We also insist that $\zeta \leq \delta_{(k)}/2$.  Finally,  
when the potential sampling locations are restricted to a finite set  of points $\{X_i, \; i = 1, \ldots, N\}$, the above constructions are modified in the obvious way, with sampling at random from  the $N$ potential locations  replacing uniform random sampling of points $x \in {\cal D}$, with the proviso that it will be impossible to construct an \textbf{ICP($n, k, \delta, \zeta$)} design for some combinations of $n$, $k$, $\delta$ and $\zeta$.

\section{Simulation study}\label{sec:simulate}
 We have carried out simulation studies of our proposed designs to illustrate the gains in predictive efficiency that can be achieved using inhibitory designs when covariance parameters have to be estimated. In our simulation studies, we evaluate our performance criterion (\Cref{eq:apv}) at the estimated parameter values using the plug-in prediction method \citep{Diggle2007}. We simulate data on the unit square [0, 1]$^2$, evaluate the integral in \Cref{eq:apv} by numerical quadrature over a 64 $\times$ 64 prediction grid, and approximate the expectation of the integral by a Monte Carlo average over $s$ = 1500 independent simulations of measurement data $Y$. We consider two model classes for the data-generation process, namely the linear Gaussian and logistic binomial geostatistical models. Both include an unobserved stationary Gaussian process $S(x)$ with mean zero, variance $\sigma^2$ = 1 and  Mat\'{e}rn correlation \citep{Matern1960}. 

In the linear Gaussian model, 

\begin{equation}\label{eq:gaussmodel}
	Y|S \sim N(\mu, \tau^2)
\end{equation}
where $\mu = S(x)$,  whilst in the logistic binomial model, 
\begin{equation}\label{eq:binmodel}
	Y|S, U \sim Bin(n, p),
\end{equation}
where $\log(p/ 1 - p) = S(x) + U$ and $U$ is Gaussian white noise with variance $\tau^2$. In both cases, the predictive target is $S$.

We used a fixed value of the correlation shape parameter, $\kappa$ = 1.5, but varied the correlation range parameter $\phi$ and the nugget variance $\tau^2$.

\subsection{Linear Gaussian Model}\label{sec:gauss-model} 
\begin{figure}
    \centering
    \begin{subfigure}[b]{0.46\textwidth}
        \centering
        \includegraphics[width=\textwidth, height = 7.5cm]{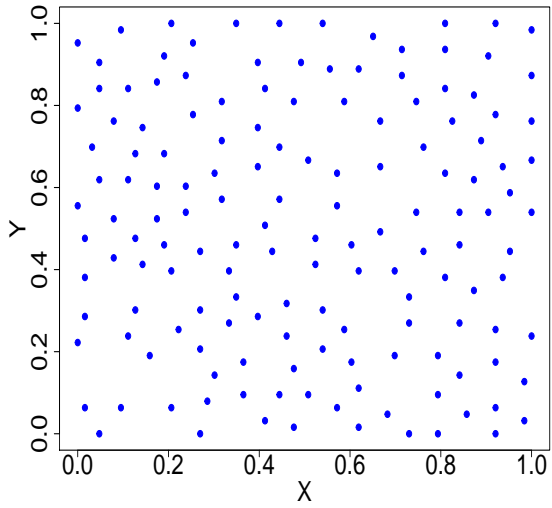}
        \caption{}
        \label{fig:InhibitoryDesign}
    \end{subfigure}
    \hfill
    \begin{subfigure}[b]{0.46\textwidth}
        \centering
        \includegraphics[width=\textwidth, height = 7.5cm]{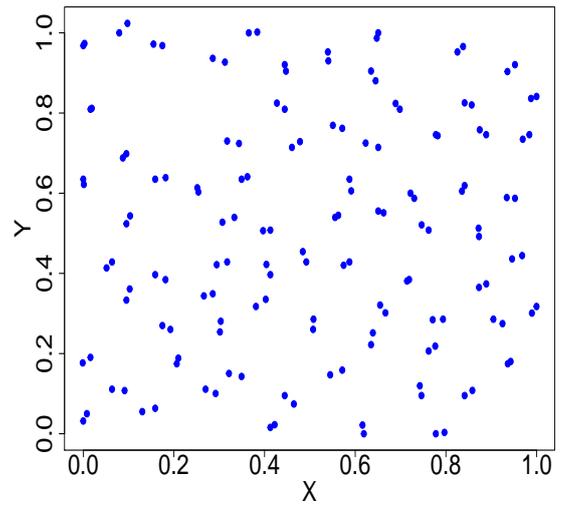}
        \caption{}
        \label{fig:InhbitoryDesignCP75}
    \end{subfigure}
		\caption{Simple inhibitory design, $\delta = 0.06$ (a). Inhibitory design with $k =  75$ close pairs, $\delta = 0.085$ for $n - k$ inhibitory design points (b). The inhibitory distance $\delta$ for (b) varies with number of close pairs $k$. Sample size $n = 150$ for each of the designs.}
	\label{fig:compare}
\end{figure}

For each parameter combination, we generated data at $n=150$ sampling locations.  \Cref{fig:InhibitoryDesign} shows an inhibitory design without close pairs and $\delta = 0.06$, corresponding to packing density $\rho \approx 0.424$, whilst \Cref{fig:InhbitoryDesignCP75} shows a design with $k = 75$ close pairs and $\delta = 0.085$, so that the $n-k=75$ inhibitory design points also have packing density 0.424.

\begin{figure}
    \centering
    \begin{subfigure}[b]{0.46\textwidth}
        \centering
        \includegraphics[width=\textwidth, height = 7.5cm]{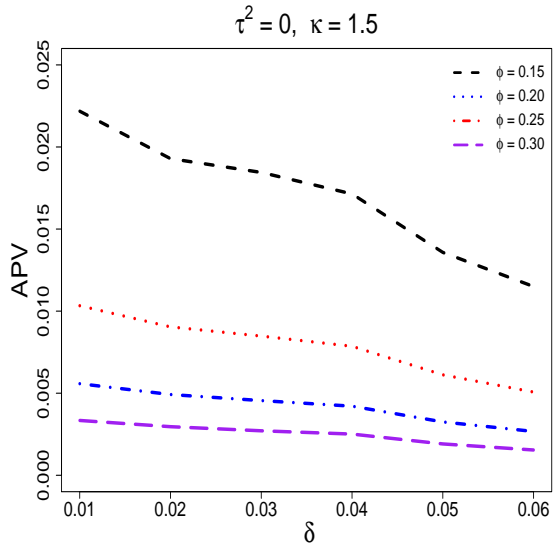}
        \caption{}
        \label{fig:IH_Delta_Gauss1}
    \end{subfigure}
    \hfill
    \begin{subfigure}[b]{0.46\textwidth}
        \centering
        \includegraphics[width=\textwidth, height = 7.5cm]{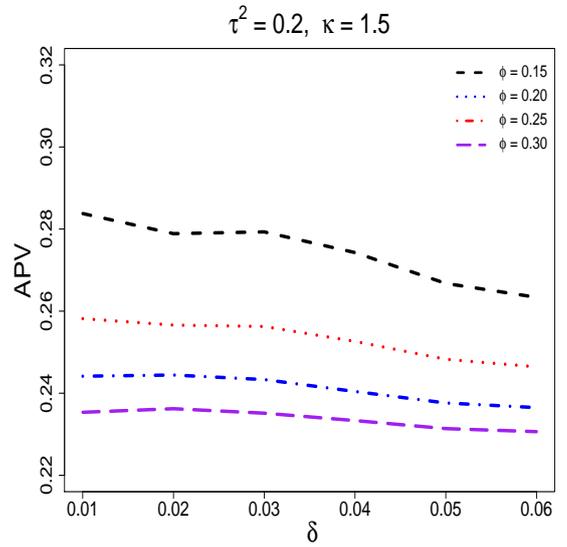}
        \caption{}
        \label{fig:IH_Delta_Gauss2}
    \end{subfigure}
		\caption{Average prediction variance for varying simple inhibitory designs, $\delta = $ 0.01 to 0.06, $\kappa = 1.5, \sigma^2 = 1$ and $n = 150$. Panel (a) $\tau^2 = 0$ and panel (b)  $\tau^2 = 0.2$.}
	\label{fig:IH_Delta_Gauss}
\end{figure}

\Cref{fig:IH_Delta_Gauss} shows the design performance as $\delta$ varies between 0.01 and 0.06, $\phi=0.15, 0.30, 0.25$ and 0.3, and for noise-to-signal ratios $\tau^2=0$ and 0.2. Results (not shown)
for  $\tau^2$ = 0.05, 0.1 and 0.4 show similar trends. These results indicate that designs with larger $\delta$ perform better, i.e.  spatial predictions become more precise with increasing regularity of the design.

\begin{figure}
	\centering
		\includegraphics[width = 17.5cm, height = 12cm]{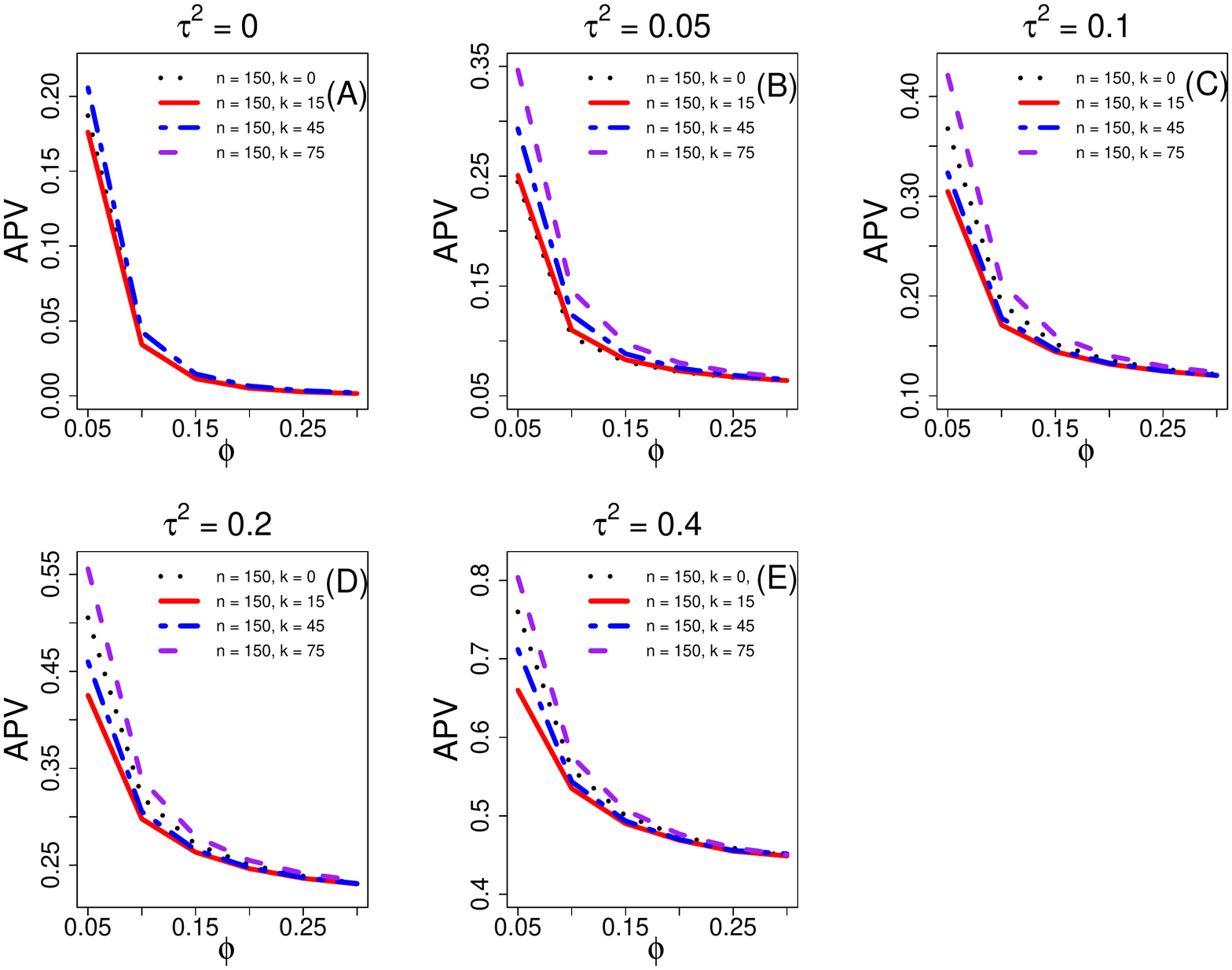}
	\caption{Comparing the efficiencies of inhibitory designs: without close pairs, with 15, 45 and 75 close pairs. The fixed total $n = 150$ for each of the designs.}
	\label{fig:CPDesignK15Tau}
\end{figure}

Our comparison of inhibitory designs with and without close pairs indicates that designs with an intermediate number of close pairs give the best performance. However, when $\tau^2$ is close to zero the benefits of close pairs are negligible, see \Cref{fig:CPDesignK15Tau} panels A -- B. In contrast, when $\tau^2$ is larger, close pairs show substantial benefit, see \Cref{fig:CPDesignK15Tau} panels C -- E.

\subsection{Binomial Model}\label{sec:binom-model}

\begin{figure}
    \centering
    \begin{subfigure}[b]{0.46\textwidth}
        \centering
        \includegraphics[width=\textwidth, height = 7.5cm]{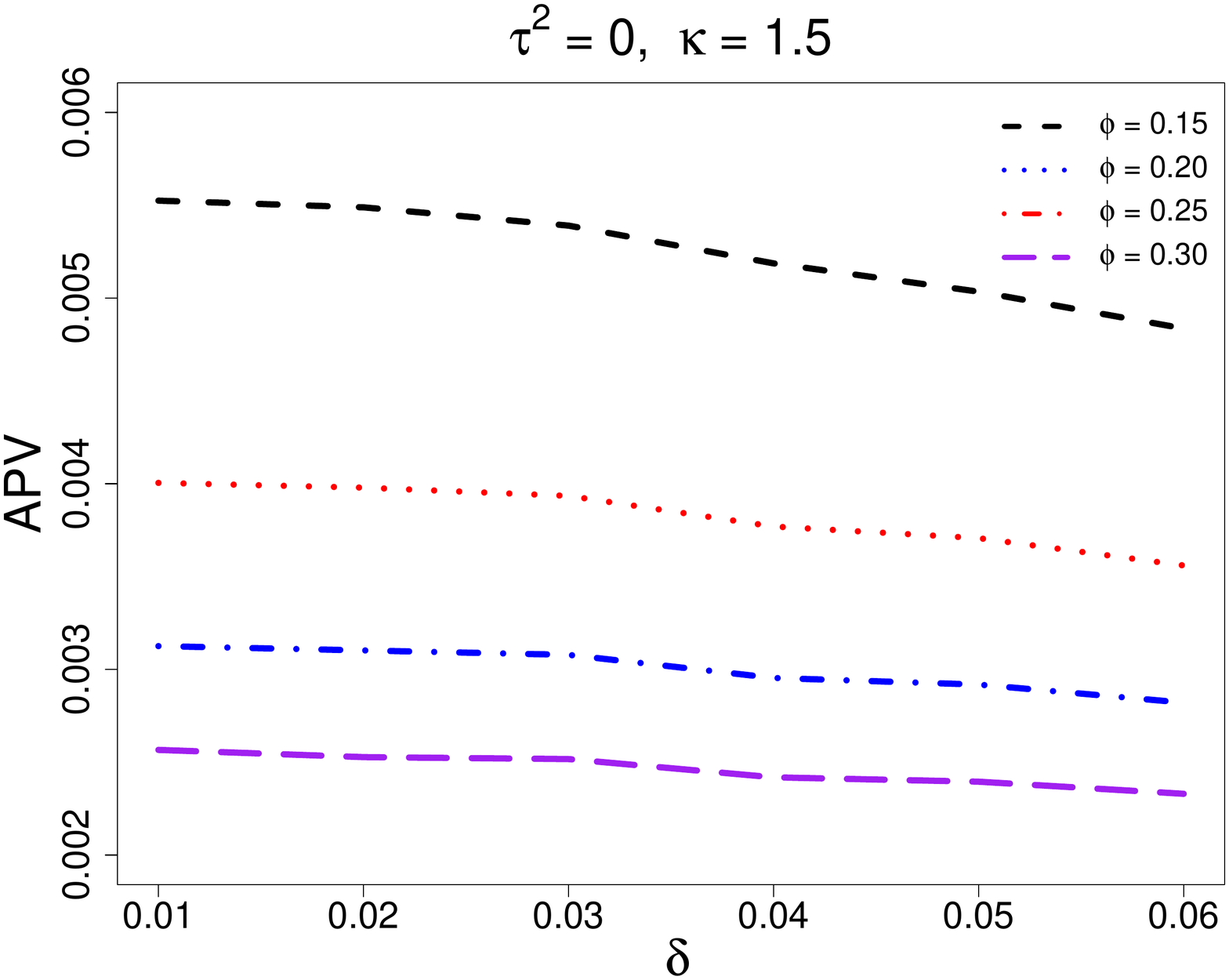}
        \caption{}
        \label{fig:IH_Delta_Binom1}
    \end{subfigure}
    \hfill
    \begin{subfigure}[b]{0.46\textwidth}
        \centering
        \includegraphics[width=\textwidth, height = 7.5cm]{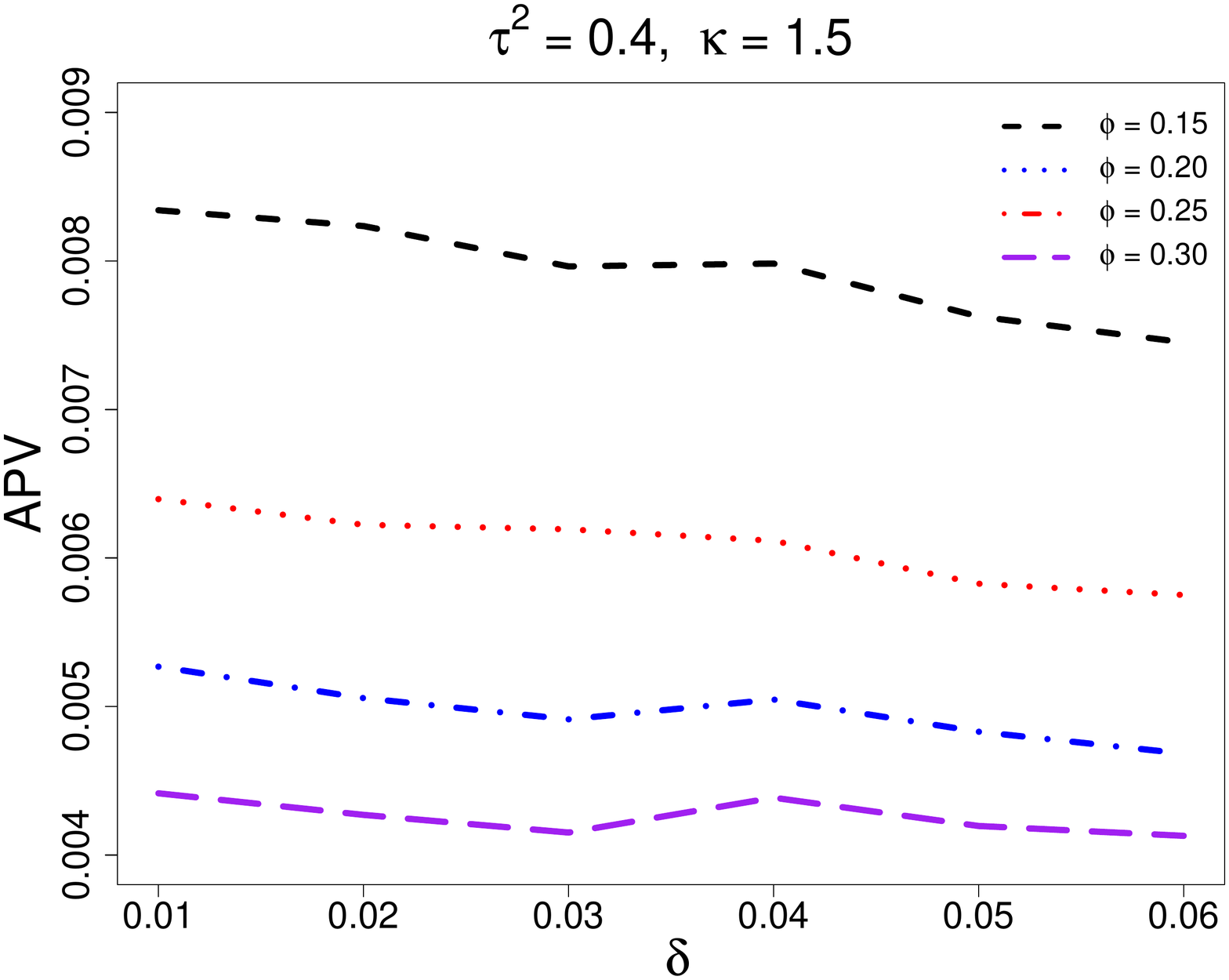}
        \caption{}
        \label{fig:IH_Delta_Binom3}
    \end{subfigure}
		\hfill
		\begin{subfigure}[b]{0.46\textwidth}
        \centering
        \includegraphics[width=\textwidth, height = 7.5cm]{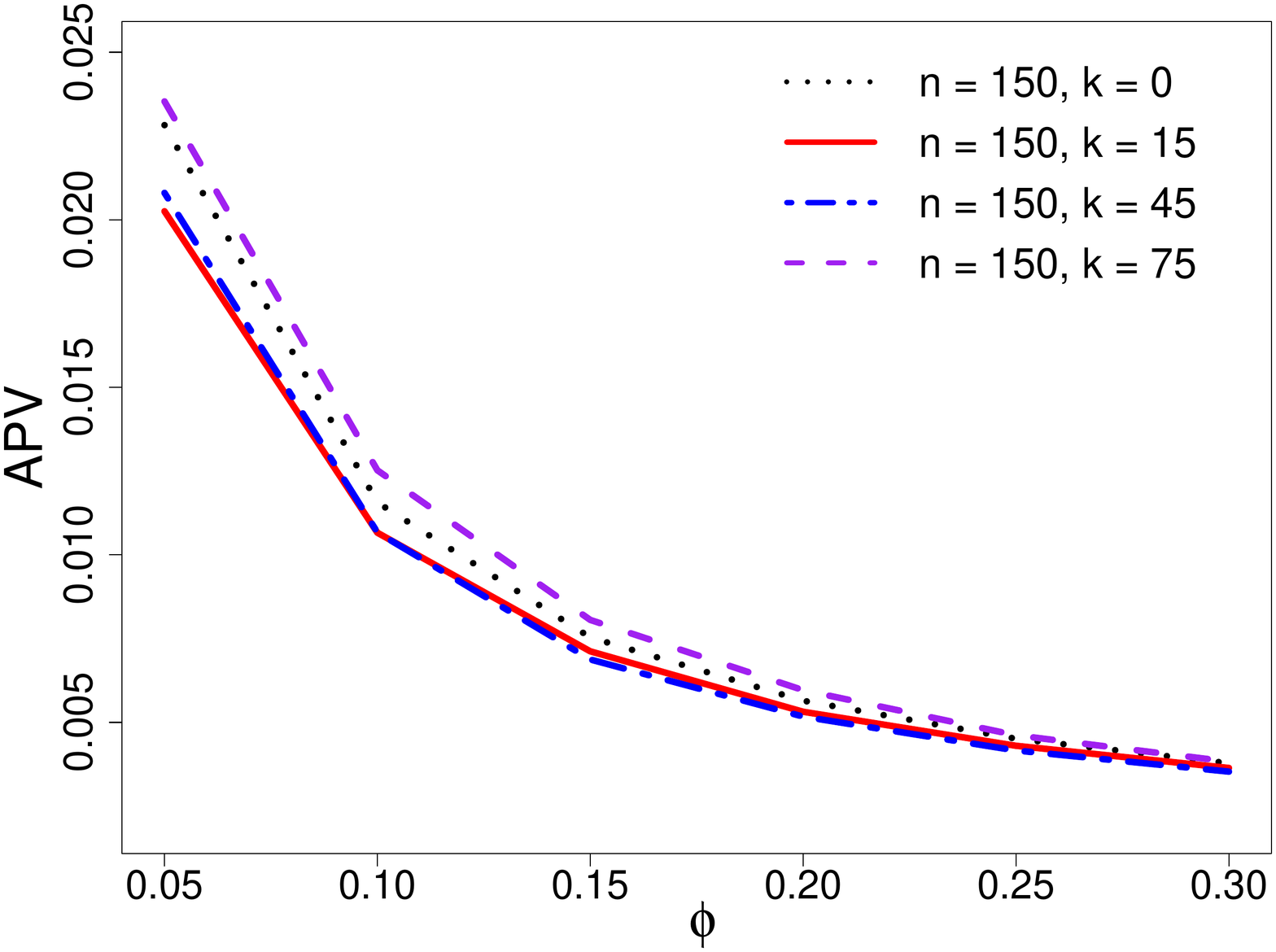}
        \caption{}
        \label{fig:CP_Design_K15Tau02_Binom}
    \end{subfigure}
		\caption{Average prediction variance for varying simple inhibitory designs - Binomial model, $\delta = $ 0.01 to 0.06, $\kappa = 1.5, \sigma^2 = 1$ and $n = 150$. Panel (a) $\tau^2 = 0$ and panel (b)  $\tau^2 = 0.4$. Panel (c) compares the efficiencies of inhibitory designs with 15, 45 and 75 close pairs. The fixed total $n = 150$ for each of the designs.}
	\label{fig:IH_Delta_Binom}
\end{figure}

We simulated binomial data-sets with 10 trials at each of $n = 150$ grid points and probabilities given by the anti-logit of the simulated values of the Gaussian process. For each combination of parameters, we approximated the expectation in \Cref{eq:apv} by a Monte Carlo average over $s$ = 1000 independent simulations of $Y.$ \Crefrange{fig:IH_Delta_Binom1}{fig:IH_Delta_Binom3} show that inhibitory designs with $\delta = 0.06$ give the best results, agreeing with the findings in \Cref{sec:gauss-model}, \Cref{fig:IH_Delta_Gauss}. Similarly,  \Cref{fig:CP_Design_K15Tau02_Binom} again
shows that inhibitory designs with an intermediate number of close pairs give the best performance when $\tau^2$ is relatively large.

\section[Application]{Application: sampling to predict spatial variation in malaria prevalence in the majete perimeter}\label{sec:applic}
In this section, we illustrate the use of our proposed inhibitory design strategy to construct a survey sample for mapping malaria prevalence in an area surrounding Majete wildlife reserve (MWR) within Chikwawa district, Malawi. The MWR is situated in the lower Shire valley at the edge of the African Rift Valley in the southern part of Malawi (15.97$^{\circ}$ S; 34.76$^{\circ}$ E). The reserve is crossed by two perennial rivers, the Shire and Mkurumadzi Rivers. Mwanza River runs near the western and southern boundaries of the park. In the wet season, there are also seasonal pools and many seasonal streams. Most rainfall occurs in the wet season, which lasts from November to April. Annually, the precipitation is 680 to 800 mm in the eastern lowlands and 700 to 1000 mm in the western highlands \citep{Wienand2013}. With an average daily temperature of 28.4 $^{\circ}$C, the wet season is slightly warmer than the dry season (average daily temperature 23.3 $^{\circ}$C), though the hottest months are September to November, at the end of the dry season \citep{Staub2013}.

\begin{figure}
	\centering
		\includegraphics[width=10cm, height = 9cm]{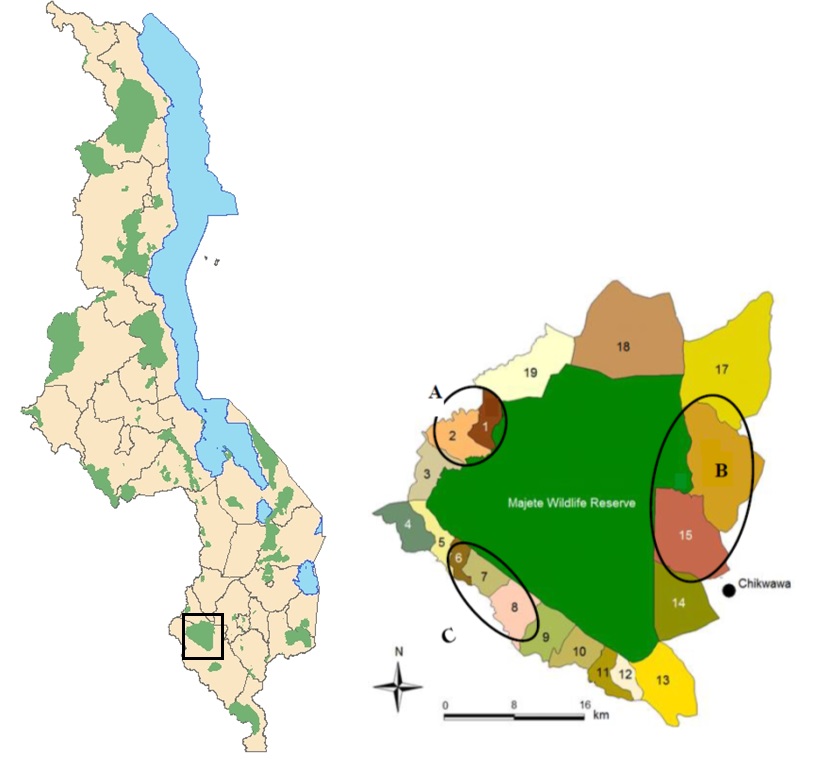}
	\caption{The map of Malawi, showing Majete Wildlife Reserve highlighted (left) and its perimeter with focal areas A, B and C highlighted (right).}
	\label{fig:MalawiMajete}
\end{figure}

The Majete malaria project (MMP) is a five-year monitoring and evaluation study of malaria prevalence, with an embedded randomised trial of community-level interventions intended to reduce malaria transmission. The study takes place in the ``Majete Perimeter'', which is the zone surrounding the MWR. The whole perimeter is home to a population of $\approx$ 100,000. \Cref{fig:MalawiMajete} shows the location of the study area, covering the unprotected zone surrounding the game park. The perimeter is subdivided into 19 community-based organizations (CBOs). Three sets of these CBOs (CBOs -- 1 \& 2, CBOs -- 6, 7 \& 8 and CBOs --15 \& 16) are administrative units, which within MMP are called {\it focal areas} A, B and C. The first stage in the geostatistical design was a complete enumeration of households in the study region, including their geo-location collected using Global Positioning System (GPS) devices on a Samsung Galaxy Tab 3 running Android 4.1 Jellybean operating system. These devices are accurate to within 5m.

\begin{figure}
	\centering
		\includegraphics[width=15cm, height = 12cm]{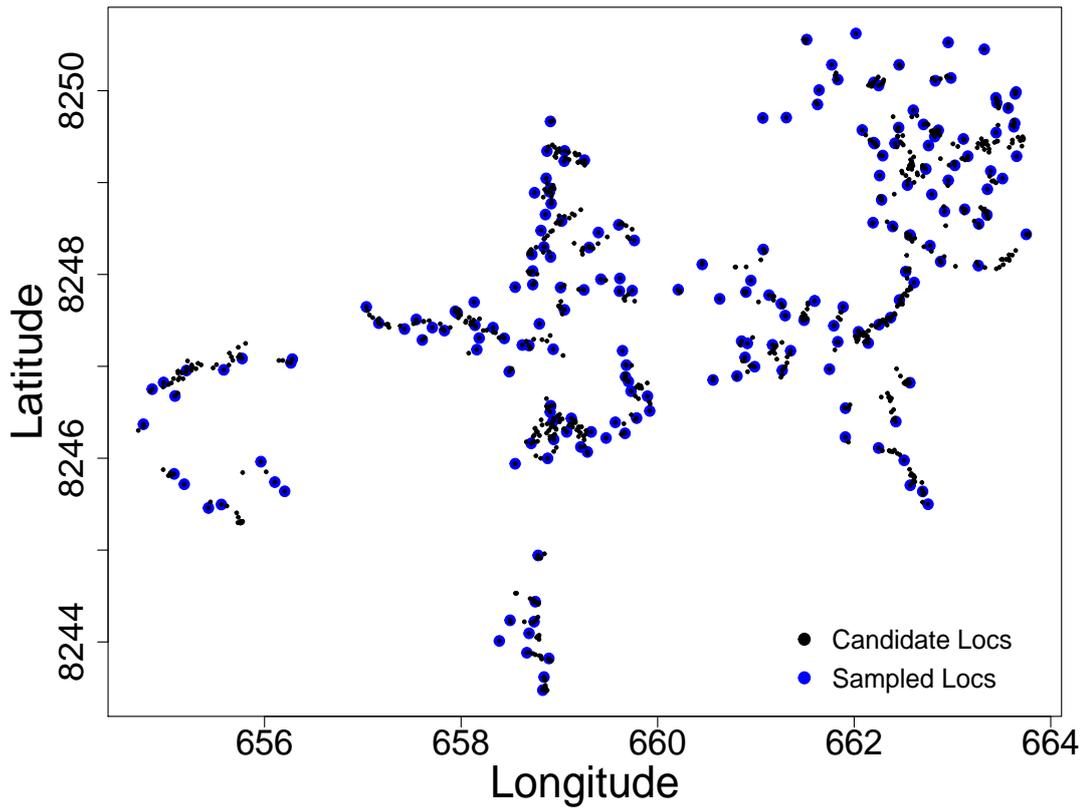}
	\caption{Inhibitory plus close pairs design locations (blue dots) and all potential sampling locations (black dots), in focal area A}
	\label{fig:FocalAreaA}
\end{figure}

The sampling unit is a household. We first fit a Binomial model, with four parameters representing the mean, two variance components and the rate of decay of spatial correlation with distance, to data from focal area B, then use the resulting estimated covariance model to inform an optimal sampling design for focal area A, whilst allowing for re-estimation of the model parameters. \Cref{tab:CovTable} shows the estimated covariance parameters. Using these estimates, the optimal design in focal area A is shown in \Cref{fig:FocalAreaA}. From a candidate set of 857 households we sample 200, of which 20 are close pairs. The blue dots represent the 200 sampled households and the black dots the locations of the other 657  households in focal area A. The optimised sampling locations provide a good spatial coverage of the study area, which is advantageous for efficient spatial prediction, whilst the inclusion of the close pairs is advantageous for parameter estimation.

\begin{table}
\centering
\begin{tabular}{llll}
\hline
\multicolumn{1}{c}{Term} & \multicolumn{1}{c}{Estimate} & \multicolumn{2}{c}{95 \% confidence interval} \\
\hline
\multicolumn{1}{c}{Intercept} & \multicolumn{1}{c}{-1.90986} & \multicolumn{1}{c}{(-2.19000,  } & \multicolumn{1}{c}{-1.62973)} \\
\multicolumn{1}{c}{$\sigma^2$} & \multicolumn{1}{c}{0.53016} & \multicolumn{1}{c}{(0.31787,  } & \multicolumn{1}{c}{0.88422)} \\
\multicolumn{1}{c}{$\tau^2$} & \multicolumn{1}{c}{0.26328} & \multicolumn{1}{c}{(0.07426,  } & \multicolumn{1}{c}{0.93341)} \\
\multicolumn{1}{c}{$\phi$} & \multicolumn{1}{c}{0.31913} & \multicolumn{1}{c}{(0.13320,  } & \multicolumn{1}{c}{0.76459)} \\
\hline
\end{tabular}
\caption{Monte Carlo maximum likelihood estimates and 95 \% confidence intervals for the covariance model fitted to malaria prevalence data in Majete focal area B.}
\label{tab:CovTable}
\end{table}

\section{Discussion}\label{sec:discuss}
Parameter values are usually unknown in practice. Designing for efficient spatial prediction with estimated parameters involves a compromise. In this paper we have proposed and demonstrated a class of inhibitory sampling designs for accurate spatial prediction with estimated covariance model parameters. The design strategies described in \Cref{sec:inhibitory} are specifically intended to deliver efficient mapping of the complete surface, $S(x)$, over the region of interest. We considered inhibitory designs with and without close pairs of sampling locations. Inhibitory designs are random designs that generate spatially regular configurations of design points. 

Our proposed designs incorporate the widely accepted concept that spatial prediction is improved by using a more-regular-than-randomly configuration of sampling locations \citep{Olea1984}. Our simulation studies show that when the same data are used for both parameter estimation and spatial prediction, the optimum inhibitory design includes a small proportion of close pairs (between 10\% 
and 30\% in our examples). This is consistent with previously expressed views that in order to compromise between prediction accuracy and efficient parameter estimation, optimal geostatistical designs should include close pairs in an otherwise spatially regular design \citep{Lark2002, Diggle2006, Muller2007}. However, our results also show that with our proposed class of designs, clear benefits for including close pairs are only realised when the nugget variance is relatively large. We conjecture that this is a consequence of the fact that inhibitory designs avoid the rigidity of lattice designs, resulting in a more varied set of inter-point distances.

We have approached the sampling design problem using inhibitory designs assuming a stochastic process with a stationary covariance structure. These are common and reasonable assumptions to make in geostatistics. However, one limitation is that when explanatory variables are available, their spatial distribution will also affect design performance. However, numerical optimisation of a performance criterion such as \Cref{eq:apv} in the presence of explanatory variables involves no additional principles.

\section*{Acknowledgements}
The MMP study was generously supported by Dioraphte Foundation, The Netherlands. The content is solely the responsibility of the authors and does not necessarily represent the official views of the funders.
	
	\section*{Funding}
	Michael Chipeta is supported by an ESRC-NWDTC Ph.D. studentship (grant number ES/J500094/1).	Dr Dianne Terlouw, Prof. Kamija Phiri and Prof. Peter Diggle are supported by the Majete integrated malaria control project grant.

\end{document}